\documentclass[12pt]{article}

\usepackage{graphicx}
\usepackage{lscape}

  \def\be{\begin{equation}}      
  \def\ee{\end{equation}}    \def\beq{\begin{eqnarray}}
  \def\dis{\displaystyle}    \def\eeq{\end{eqnarray}}
       \def\m{\multicolumn}

\begin{document}

\begin{center}
{\large \textbf{Decay Properties of $D$ and $D_s$
mesons}}\\
\textbf{\footnote{Email: azadpatel2003@gmail.com}Bhavin Patel and {$^*$}P C Vinodkumar}

 LDRP- Institute of Technology and Research, Gandhinagar- 382 015, Gujarat, INDIA.\\ 
 {$^*$}Department of Physics, Sardar Patel University, Vallabh Vidyanagar- 388 120, Gujarat, INDIA
\end{center}
\begin{abstract}
The decay rates and spectroscopy of the $D$ and $D_s$ mesons are
computed in a nonrelativistic phenomenological quark-antiquark
potential of the type $V(r)=-\frac{4}{3}\frac{\alpha_s}{r}+A
r^{\nu}$ with different choices of $\nu$. Numerical method to solve
the Schr\"{o}dinger equation has been used to obtain the
spectroscopy of $q\bar{Q}$ mesons. The spin hyperfine, spin-orbit
and tensor components of the one gluon exchange interactions  are
employed to compute the spectroscopy of the excited $S$ states, low
lying $P$-waves and $D$-waves. The numerically obtained radial
solutions are employed to obtain the decay constant and leptonic
decay widths. It has been observed that predictions of the
spectroscopy and the decay widths are consistent with other model
predictions as well as with the known experimental values.
\end{abstract}
\section{Introduction}
Spectroscopy of hadrons containing heavy flavours has attracted
considerable interest in recent years due to many experimental
facilities such as the BES at the Beijing Electron Positron Collider
(BEPC), E835 at Fermilab, and CLEO at the Cornell Electron Storage
Ring (CESR) $\emph{etc.}$, worldover. They have been able to collect
huge data samples in the heavy flavour sector. Where as B-meson
factories, BaBar at PEP-II and Belle at KEKB are working on the
observation of new and possibly exotic hadronic states. All these
experiments are capable of discovering new hadrons, new production
mechanisms, new decays and transitions and in general will be
providing  high precision data sample with better stastics and
higher confidence level. After having played a major role in the
foundation of QCD, heavy hadron spectroscopy has witnessed in the
last few years a renewal of interest led by the many new data coming
from the B factories, CLEO and the Tevatron and by the progress made
in the theoretical methods. The remarkable progress at the
experimental side, with various high energy machines such as BaBar,
BELLE, B-factories, Tevatron, ARGUS collaborations, CLEO, CDF,
D\O~\emph{etc.,} for the study of hadrons has opened up new
challenges in the theoretical understanding of light-heavy flavour
hadrons. The existing results on excited heavy-light mesons are
therefore partially inconclusive, and even contradictory in several
cases. The predictions of masses of heavy-light system for ground
state as well as excited state are few from the theory
\cite{Godfrey1991,Pierro2001,ebert1998,Bardeen2003,Colangelo2003,Falk2003,Eichten2003}.
In the open charm sector, the observation of a charm-strange state,
the $D_{sJ}^*(2317)$  state \cite{Bab03} by BaBar Collaboration. It
was confirmed by CLEO Collaboration at the Cornell Electron Storage
Ring \cite{Cle03} and also by Belle Collaboration at KEK
\cite{Bel04}. Besides, BaBar had also pointed out to the existence
of another charm-strange meson, the $D_{sJ}(2460)$ \cite{Bab03}.
This resonance was measured by CLEO \cite{Cle03} and confirmed by
Belle \cite{Bel04}. Belle results \cite{Bel04} are consistent with
the spin-parity assignments of $J^P=0^+$ for the $D^*_{sJ}(2317)$
and $J^P=1^+$ for the $D_{sJ}(2460)$. Thus, these states are well
established and confirmed independently by different experiments.
They present unexpected properties, quite different from those
predicted by quark potential models. If they would correspond to
standard $P-$wave mesons made of a charm quark and a strange
antiquark their masses would be larger \cite{God91}, around 2.48 GeV
for the $D_{sJ}^*(2317)$ and 2.55 GeV for the $D_{sJ}(2460)$. They
would be therefore above the $DK$ and $D^*K$ thresholds,
respectively with being broad resonances. However the states
observed by BaBar and CLEO are very narrow, $\Gamma < 4.6$ MeV for
the $D_{sJ}^*(2317)$ and $\Gamma < 5.5$ MeV for the $D_{sJ}(2460)$.
\\
In near future, even larger data samples are expected from the
BES-III upgraded experiments, while the B factories and the Fermilab
Tevatron will continue to supply valuable data for few more years.
Later on, the LHC experiments at CERN, Panda at GSI \emph{etc.},
will be accumulating large data sets which will offer greater
opportunities and challenges particularly in the field of heavy
flavour
physics \cite{Brambilla2007}.\\
At the hadronic scale the nonperturbative effects connected with
complicated structure of QCD vacuum necessarily  play an important
role. But our limited knowledge about the nonperturbative QCD leads
to a theoretical uncertainty in the quark- antiquark potential at
large and intermediate distances \cite{Brambilla2002}. So a
successful theoretical model can provide important information about
the quark-antiquark interactions and the behavior of QCD at the
hadronic scale. Though there exist many potential models with
relativistic and nonrelativistic considerations employed to study
the hadron properties based on its quark structure
\cite{AjayRai2008,Ajay2008,JNPandya2001,Radford2007,BuchmullerTye1981,Martin1980,Amartin1979,Quiggrosner1977,QuiggRosner1979,Eichten1978,vijayakumar2004,Altarelli1982,Ebert2003,Lakhina2006,Choi2007},
the most commonly used potential is the coulomb plus linear power
potential, $V(r)=-\frac{4}{3}\frac{\alpha_s}{r} +\sigma r$, with the
string tension $\sigma$ \cite{Gunnar2000,Alexandrou2003}. However,
for the higher excited mesonic states it is argued that the string
tension $\sigma$ must depend on the $Q\bar{Q}$ separation
\cite{Badalian2002,Badalian2008}. This corresponds to flattening of
the confinement potential at larger $r$ $(r\geq1 fm)$.  More over
the analysis based on Regge trajectories for meson states suggests
the confinement  part of the potential to have the power
$\frac{2}{3}$ instead of 1 \cite{Albertus2005,Fabre1988}.
 This has prompted us to choose a power form for the confining part
 of the interquark potential and study the properties of heavy flavour systems
 by varying the power of the confinement part of the interquark potential
 different from  1.0. \\
  Apart from the spectroscopic predictions of higher orbital states,
  other problems associated with the phenomenological models employed for the properties of mesons are
  the right predictions of their decay properties.
 For better predictions of the decay widths, many models need to incorporate
 additional terms such as the radiative contributions, higher order QCD corrections $\emph{etc.}$,
 to the conventional decay
  formula \cite{AjayRai2008,Ajay2008,EbertMod6012003,Lansberg2008,Kim2005}.\\
 The decay widths can provide an account of the compactness of
the meson system in terms of the radial wave function
 which is an useful information complementary to
spectroscopy \cite{Rosner2006}.  Other unresolved issues are related
to the hyperfine and fine structure splitting of the mesonic states
  and their intricate dependence with the constituent
  quark masses and the running strong coupling constant. Thus, in this paper we make an
attempt to study  properties like mass spectrum, decay constants and
other decay properties of the open charm mesons ($D, D_s$). We
investigate the heavy-light mass spectra of $D (c \bar q)$ and
$D_{s} (c \bar s)$ mesons in the frame work of the nonrelativistic
CPP$_\nu$ potential model. In the present study, we consider
different choices of the potential power index
$\nu$ in the range $0.1<\nu<2.0$. 
\section{Nonrelativistic Treatment for Heavy Flavour Mesons using
CPP$_\nu$} \label{NRQM_model} In general, properties of heavy
flavour mesons have been studied based on potential models in the
frame work of relativistic as well as nonrelativistic quantum
mechanics. In the limit of heavy quark mass $m_Q$ $\rightarrow$
$\infty$, heavy meson properties are governed by the dynamics of the
light quark. As such, these states become hydrogen like atoms of
hadron physics. Moreover, both the non-relativistic predictions are
in fair agreements with each other as well as with the available
experimental and lattice results. Hence, for the present study of
charm meson bound states, we consider a nonrelativistic Hamiltonian
given by \cite{AjayRai2008,Ajay2008,RaiPhD2005,Bhavin2009,
AKRai2002,AKRai2006}
\begin{equation}\label{eq:nlham}
 H= M + \frac{p^2}{2M_1} +  V(r) ,\end{equation}
here,
 \begin{equation}
 M=m_1 + m_{2} , \  \ \  \ and \   \  \  \
M_1=\frac{m_1 \ m_{2}}{m_1 + m_{2}},
\end{equation} \label{eq:mmHH}
The relative momentum of each quark is represented by $p$ and $V(r)$
is the quark-antiquark potential. Nonrelativistically, this
interaction potential consists of a central term $V_{c}(r)$ and a
spin dependent part $V_{SD}(r)$. The central part $V_{c}(r)$ is
expressed in terms of a
 vector (Coulomb)  plus a scalar (confining)  part given by
 \begin{equation} \label{eq:pote}
 V_{c}(r)=V_{V}+V_{S}=-\frac{4}{3}\frac{\alpha_s}{r} + A r^\nu
 \end{equation}
as the static quark-antiquark interaction potential
\cite{AjayRai2008}.
\begin{figure}[h]
\begin{center}
\includegraphics [height=4.0in,width=5.0in]{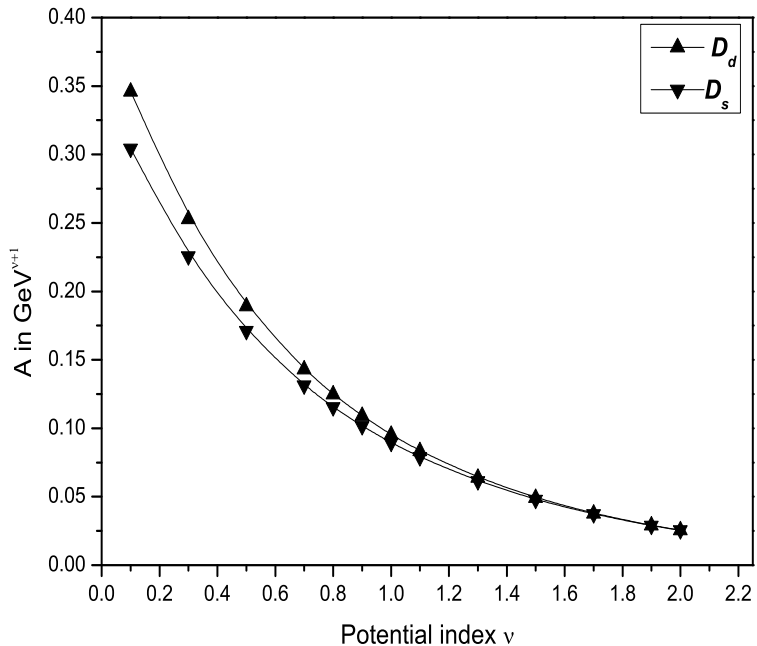}
\vspace{-0.2in} \vspace{-0.2in} \caption{Behavior of $A$ with the
potential index $\nu$ for $D$ and $D_s$ mesons. }\label{fig:R1s_LH}
\end{center}
\end{figure}
\subsection{Spin-dependent Forces}
In general, the quark-antiquark bound states are represented by
$n^{2S+1} L_J$,  identified with the $J^{PC}$ values, with $ \vec
J=\vec L + \vec S$, $ \vec S=\vec S_q + \vec S_{\bar Q}$, parity
$P=(-1)^{L+1}$ and the charge conjugation $C = (-1)^{L+S}$ with $(n,
L)$ being the radial quantum numbers. So the $S$-wave $(L=0)$ bound
states are represented by $J^{PC}=0^{-+}$ and $1^{--}$ respectively.
The $P$-wave $(L=1)$ states are represented by $J^{PC}=1^{+-}$ with
$L=1$ and $S=0$ while $J^{PC}=0^{++},\ 1^{++}$ and $2^{++}$
correspond to $L=1$ and $S=1$ respectively. Thus, the spin-spin
interaction among the constituent quarks provides the mass splitting
of $J=0^{-+}$ and $1^{--}$ states, while the spin-orbit interaction
provides the mass splitting of $J^{PC}=0^{++},\ 1^{++}$ and $2^{++}$
states. The $J^{PC} = 1^{+-}$ state with $L=1$ and $S=0$ represents
the spin average mass of the $P$-state as its spin-orbit
contribution becomes zero, while the two $J=1^{+-}$ singlet and  the
$J=1^{++}$ of the triplet $P$-states can form a mixed state. The
$D$-wave $(L=2)$ states are represented by $J^{PC}=2^{-+}$ with
$L=2$ and $S=0$ while $J^{PC}=3^{--},\ 2^{--}$ and $1^{--}$
correspond to $L=2$ and $S=1$ respectively. The $F$-wave $(L=3)$
states are represented by $J^{PC}=3^{+-}$ with $L=3$ and $S=0$ while
$J^{PC}=4^{++},\ 3^{+-}$
and $2^{++}$ correspond to $L=3$ and $S=1$ respectively. \\
\\For computing the mass difference between different degenerate
meson states, we consider the spin dependent part of the usual one
gluon exchange potential (OGEP) given by
 \cite{Lakhina2006,Branes2005,Voloshin2008,Eichten2008,Gerstein1995}.
 Accordingly, the spin-dependent part, $V_{SD}(r)$ contains
three components of the interaction terms, such as the spin-spin,
the spin-orbit and the tensor part given by \cite{Voloshin2008}
\begin{eqnarray}\label{spin}
                V_{SD}(r) &=&  V_{SS}(r)\left[S(S+1)-\frac{3}{2}\right]+
                          V_{LS}(r)\left(\vec{L}\cdot\vec{S}\right)+\cr
               && V_{T}(r) \left[S
(S+1)-\frac{3(\vec{S}\cdot\vec{r})(\vec{S}\cdot\vec{r})}{r^2}\right]
              \end{eqnarray}
The spin-orbit term containing $V_{LS}(r)$ and the tensor
 term containing $V_{T}(r)$ describe the fine structure of the meson states, while
the spin-spin term containing $V_{SS}(r)$ proportional to
$2(\vec{s_{q}}\cdot\vec{s_{\bar q}})=S(S+1)-\frac{3}{2} $ gives the
spin singlet-triplet hyperfine splitting.\\
\\The coefficient of these spin-dependent terms of Eqn.\ref{spin} can
be written in terms of the vector and scalar parts of the static
potential, $V_{c}(r)$ as \cite{Voloshin2008}
\begin{equation} \label{LS}
V_{LS}(r)=\frac{1}{2\ m_{1} m_{2}\
r}\left(3\frac{dV_{V}}{dr}-\frac{dV_{S}}{dr}\right)
\end{equation}
\begin{equation} \label{T}
V_{T}(r)=\frac{1}{6\ m_{1} m_{2}
}\left(3\frac{d^{2}V_{V}}{dr^{2}}-\frac{1}{r}\frac{dV_{V}}{dr}\right)
\end{equation}
\begin{equation} \label{SS}
V_{SS}(r)=\frac{1}{3\ m_{1} m_{2}}\nabla^{2} V_{V}=\frac{16 \ \pi
\alpha_{s}}{9\ m_{1} m_{2} } \delta^{(3)}(\vec{r})
\end{equation}
 The present study with the choices of $\nu$ in the range
$0.1<\nu<2.0,$ is  an attempt to know the predictability of the
hadron spectroscopy  with  a chosen value of mass
 parameters $(m_{1}, m_2)$  and confinements strength represented by the potential parameter $A$.
 The running strong
coupling constant appeared in the potential $V(r)$  in turn is
related to the quark mass parameter as
 \begin{equation}  \label{alpha}
 \alpha _s(\mu^2)=\frac{4 \pi}{(11-\frac{2}{3}n_{f})\ {\rm ln}(\mu^2/ \Lambda^2)}
 \end{equation}
Where, $n_f$ is the number of flavors,  $\mu$ is renormalization
scale related to the constituent quark masses as $\mu=2
m_{1}m_{2}$/$(m_{1}+m_{2})$ and $\Lambda$ is the QCD scale which is
taken as 0.150 GeV by fixing $\alpha_{s}=0.118$ at the
$Z-$boson mass (91 Ge$V$)\cite{PDG2008}.\\
\\The potential parameter, $A$ of Eqn.\ref{eq:pote} is similar to the
string strength $\sigma$ of the Cornell potential. The different
choices of $\nu$ here then correspond to different potential forms.
So, the potential parameter $A$ expressed in GeV$^{\nu+1}$ can be
different for each choices of $\nu$. The model potential parameter
$A$ and the mass parameter of the quark/antiquark ($m_{1},m_2$) are
fixed using the known ground state center of weight (spin average)
mass and the hyperfine splitting ($M_{^{3}S_{1}}-M_{^{1}S_{0}}$) of
$D$ and $D_s$ systems respectively. The spin average mass for the
ground state is computed for the different choices of $\nu$ in the
range, $0.1 \leq \nu \leq 2.0$. The spin average or the center of
weight mass, $M_{CW}$ is calculated from the known
experimental/theoretical values of the pseudoscalar ($J=0$) and
vector ($J=1$) meson states as
 \begin{equation}
M_{n, CW}=\frac{\sum\limits_{J} (2J+1)\ M_{nJ}}{\sum\limits_{J}
(2J+1)}\end{equation}
\begin{table}
\begin{center}\caption{Square of the radial wave functions at the origin($|R_{n}(0)|^2$
(in GeV$^3$))  of $Q\bar{q}$ systems in CPP$_{\nu}$.}
\label{tab:LMR01}
\begin{tabular}{rlllllll}
\hline
  \hline &\m{3}{c}{$D$}&&\m{3}{c}{$D_s$}\\
\cline{2-4}\cline{6-8}
\m{0}{c}{Model}&\m{0}{c}{1S}&\m{0}{c}{2S}&\m{0}{c}{3S}&&\m{0}{c}{1S}&\m{0}{c}{2S}&\m{0}{c}{3S} \\
  \hline
 CPP$_\nu=$ 0.1 &   0.050   &   0.011   &   0.005   &   &   0.063   &   0.014   &   0.007   \\
0.3 &   0.083   &   0.028   &   0.016   &   &   0.104   &   0.034   &   0.020   \\
0.5 &   0.111   &   0.047   &   0.031   &   &   0.139   &   0.058   &   0.038   \\
0.7 &   0.135   &   0.068   &   0.049   &   &   0.168   &   0.084   &   0.061   \\
0.8 &   0.145   &   0.079   &   0.060   &   &   0.181   &   0.098   &   0.074   \\
0.9 &   0.155   &   0.091   &   0.071   &   &   0.193   &   0.112   &   0.088   \\
1.0 &   0.164   &   0.102   &   0.083   &   &   0.204   &   0.127   &   0.103   \\
1.1 &   0.172   &   0.114   &   0.096   &   &   0.214   &   0.142   &   0.118   \\
1.3 &   0.186   &   0.139   &   0.123   &   &   0.232   &   0.172   &   0.153   \\
1.5 &   0.199   &   0.164   &   0.154   &   &   0.247   &   0.203   &   0.190   \\
2.0 &   0.222   &   0.227   &   0.237   &   &   0.276   &   0.280   &   0.292   \\
\hline \hline
\end{tabular}
\end{center}
\end{table}

\begin{table}
\begin{center}\caption{The $\ell^{th}$ derivative of orbitally excited radial wave functions at the origin ($|R_{n}^{\ell}(0)|$ in GeV$^{(\frac{3}{2}+\ell)}$) of $Q\bar{q}$ systems in  CPP$_{\nu}$.} \label{tab:LMR02}
\begin{tabular}{rlllllllll}
\hline

  \hline &\m{4}{c}{$D$}&&\m{4}{c}{$D_s$}\\
\cline{2-5}\cline{7-10}
\m{0}{c}{Model}&\m{0}{c}{1P}&\m{0}{c}{2P}&\m{0}{c}{1D}&\m{0}{c}{1F}&&\m{0}{c}{1P}&\m{0}{c}{2P}&\m{0}{c}{1D}&\m{0}{c}{1F} \\
  \hline
 CPP$_\nu=$ 0.1 &   0.0110  &   0.0083  &   0.0008  &   0.0001  &   &   0.0130  &   0.0097  &   0.0010  &   0.0001  \\
0.3 &   0.0253  &   0.0210  &   0.0034  &   0.0006  &   &   0.0299  &   0.0253  &   0.0044  &   0.0008  \\
0.5 &   0.0392  &   0.0358  &   0.0076  &   0.0018  &   &   0.0474  &   0.0428  &   0.0097  &   0.0024  \\
0.7 &   0.0551  &   0.0520  &   0.0129  &   0.0040  &   &   0.0634  &   0.0619  &   0.0165  &   0.0055  \\
0.8 &   0.0603  &   0.0604  &   0.0159  &   0.0055  &   &   0.0711  &   0.0722  &   0.0203  &   0.0076  \\
0.9 &   0.0666  &   0.0692  &   0.0192  &   0.0073  &   &   0.0793  &   0.0826  &   0.0245  &   0.0103  \\
1.0 &   0.0732  &   0.0780  &   0.0226  &   0.0093  &   &   0.0881  &   0.0931  &   0.0288  &   0.0126  \\
1.1 &   0.0790  &   0.0870  &   0.0261  &   0.0115  &   &   0.0934  &   0.1029  &   0.0335  &   0.0157  \\
1.3 &   0.0916  &   0.1043  &   0.0334  &   0.0166  &   &   0.1088  &   0.1243  &   0.0426  &   0.0227  \\
1.5 &   0.1022  &   0.1221  &   0.0408  &   0.0225  &   &   0.1189  &   0.1459  &   0.0521  &   0.0307  \\
2.0 &   0.1234  &   0.1653  &   0.0593  &   0.0397  &   &   0.1457  &   0.1969  &   0.0756  &   0.0542  \\
\hline \hline
\end{tabular}
\end{center}
\end{table}

 \subsection{Spectra of Heavy - Light Flavour
($Q \bar q$) Mesons}
 The spectra of the heavy-light mesons are
calculated using nonrelativistic hamiltonian as given by
Eqn.\ref{eq:nlham}, where $m_1=m_{Q}$ and $m_2=m_{\bar q}$.
 The spin average masses of $D^*-D$ and the $D_s^*-D_{s}$ mesons are computed using the
experimental values of $M_{D}=$ 1.869 GeV, $M_{D^*}=$2.010 GeV,
$M_{D_s}=$1.968 GeV, $M_{D_s^*}=$ 2.112 GeV respectively
\cite{PDG2008}.  \\
We employ the numerical approach as given by \cite{Lucha1999} to
find the eigen values and radial wave functions of the respective
Schr\"{o}dinger equation. The potential parameter $A$, is made to
vary with $\nu$, keeping the quark mass parameter fixed for each
choices of $Q \bar q$ system. It is observed that the hyperfine
splitting of the $1 ^3S_{1}$ and $1 ^1S_{0}$ states are very
sensitive to the choices of $m_Q$ and $A$. The most suitable values
of the quark mass parameter are found to be $m_c=1.28$ GeV,
$m_d=0.35$ GeV $m_s=0.500$ GeV for our present study. The
corresponding $A$ values obtained from the $1S$ fitting and are
plotted in Fig.\ref{fig:R1s_LH} against the potential index $\nu$  of the
$D$ and $D_s$ systems. Just like the string tension $\sigma(r)$ of
the Cornell potential was made to vary for excited states
\cite{Badalian2002,Badalian2008}, we allow $A$ to vary mildly with
radial quantum number $n$
 as $A \rightarrow \frac{A}{(n)^{\frac{1}{4}}}$ for computing the spin independent masses of
the orbital excited  $(nL)$ states. The variation in $A$ can be
justified by similar arguments for the changes in $\alpha_{s}$ with
the average kinetic energy. Here, as the system get excited, the
average kinetic energy increases and hence the potential strength
(the spring tension) reduces. With this mild state dependence on the
potential parameter $A$, we obtain the spin average masses of the
orbital excited states closer to the experimentally known $D$  and
$D_s$ systems. The computed values of the radial wave function at
the origin $|R_{n\ell}^{\ell}(0)|$ for different
 states are listed in Table \ref{tab:LMR01} ($nS-$states) and
Table \ref{tab:LMR02} ($1P, 2P, 1D, 1F-$states) for all the $c \bar
q$ ($Q \ \epsilon \ c$ and $q \ \epsilon \ u/d, s)  $
combinations. Using the spin dependent potential given by
Eqn.\ref{spin}, we compute the masses of the different $n^{2 S+1}
L_{J}$ low lying states of  $c \bar q$  and are listed in Table
\ref{tab:LM03} and \ref{tab:LM04} in the case of $D$ and $D_s$
mesons respectively. The available experimental values as well as
other model predictions are
also listed for comparison.\\
\begin{landscape}
\begin{table}[h]
\begin{center}
\caption{Mass spectra (in GeV) of $D$ meson.}\label{tab:LM03}
\begin{tabular}{lllllllllllllll}
\hline \hline State&\multicolumn{6}{r}{\textbf{
Potential index  $\nu$}}&&&&&Expt.&RQM&RQM&BSU \\
\cline{2-11}
&0.1&0.7&0.8&0.9&1.0&1.1&1.3&1.5&1.7&2.0&\cite{PDG2008}&\cite{Ebert2003}&\cite{Pierro2001}&\cite{Lahde2000}  \\
\hline
$1 ^3S_1$   &   1.985   &   2.007   &   2.010   &   2.013   &   2.015   &   2.018   &   2.021   &   2.025   &   2.028   &   2.031   &   2.010   &   2.009   &   2.005   &   2.006   \\
$1^1S_0$    &   1.932   &   1.864   &   1.855   &   1.848   &   1.841   &   1.834   &   1.823   &   1.813   &   1.805   &   1.794   &   1.869   &   1.875   &   1.868   &   1.874   \\
\vspace{-0.1in}\\
$1 ^3P_2$   &   2.070   &   2.268   &   2.294   &   2.319   &   2.342   &   2.364   &   2.404   &   2.440   &   2.472   &   2.514   &   2.460   &   2.459   &   2.460   &   2.477   \\
$1 ^3P_1$   &   2.072   &   2.282   &   2.310   &   2.336   &   2.361   &   2.384   &   2.426   &   2.465   &   2.498   &   2.542   &       &   2.414   &   2.417   &   2.407   \\
$1 ^3P_0$   &   2.068   &   2.261   &   2.287   &   2.312   &   2.335   &   2.357   &   2.398   &   2.434   &   2.467   &   2.510   &       &   2.438   &   2.490   &   2.341   \\
$1 ^1P_1$   &   2.062   &   2.216   &   2.235   &   2.253   &   2.269   &   2.285   &   2.312   &   2.337   &   2.358   &   2.385   &       &   2.501   &   2.377   &   2.389   \\
\vspace{-0.1in}\\
$2 ^3S_1$   &   2.011   &   2.303   &   2.350   &   2.398   &   2.445   &   2.491   &   2.582   &   2.668   &   2.751   &   2.868   &       &   2.629   &   2.692   &   2.601   \\
$2^1S_0$    &   1.998   &   2.226   &   2.261   &   2.296   &   2.329   &   2.362   &   2.425   &   2.483   &   2.538   &   2.612   &       &   2.579   &   2.589   &   2.540   \\
\vspace{-0.1in}\\
$1 ^3D_3$   &   2.106   &   2.456   &   2.508   &   2.558   &   2.605   &   2.651   &   2.736   &   2.816   &   2.887   &   2.984   &       &       &   2.799   &   2.688   \\
$1 ^3D_2$   &   2.106   &   2.454   &   2.504   &   2.552   &   2.597   &   2.639   &   2.717   &   2.788   &   2.851   &   2.933   &       &       &   2.775   &   2.727   \\
$1 ^3D_1$   &   2.105   &   2.459   &   2.512   &   2.564   &   2.613   &   2.661   &   2.751   &   2.836   &   2.913   &   3.019   &       &       &   2.833   &   2.750   \\
$1 ^1D_2$   &   2.104   &   2.455   &   2.509   &   2.561   &   2.612   &   2.662   &   2.756   &   2.846   &   2.930   &   3.046   &       &       &   2.795   &   2.689   \\
\vspace{-0.1in}\\
$2 ^3P_2$   &   2.037   &   2.432   &   2.499   &   2.565   &   2.631   &   2.696   &   2.824   &   2.949   &   3.069   &   3.241   &       &       &   3.035   &   2.860   \\
$2 ^3P_1$   &   2.038   &   2.443   &   2.511   &   2.580   &   2.648   &   2.715   &   2.847   &   2.976   &   3.100   &   3.279   &       &       &   2.995   &   2.802   \\
$2 ^3P_0$   &   2.036   &   2.427   &   2.493   &   2.559   &   2.624   &   2.689   &   2.816   &   2.940   &   3.059   &   3.231   &       &       &   3.045   &   2.758   \\
$2 ^1P_1$   &   2.033   &   2.394   &   2.454   &   2.512   &   2.570   &   2.627   &   2.736   &   2.843   &   2.944   &   3.087   &       &       &   2.949   &   2.792   \\
\vspace{-0.1in}\\
$3 ^3S_1$   &   2.005   &   2.468   &   2.553   &   2.640   &   2.727   &   2.816   &   2.994   &   3.172   &   3.348   &   3.607   &       &       &   3.226   &   2.947   \\
$3^1S_0$    &   1.999   &   2.413   &   2.486   &   2.560   &   2.634   &   2.708   &   2.855   &   2.999   &   3.139   &   3.340   &       &       &   3.141   &   2.904   \\
\vspace{-0.1in}\\
$1 ^3F_4$   &   2.127   &   2.607   &   2.683   &   2.757   &   2.829   &   2.900   &   3.033   &   3.159   &   3.276   &   3.436   &       &       &   3.091   &       \\
$1 ^3F_3$   &   2.127   &   2.599   &   2.671   &   2.741   &   2.807   &   2.871   &   2.989   &   3.096   &   3.192   &   3.318   &       &       &   3.074   &       \\
$1 ^3F_2$   &   2.127   &   2.611   &   2.688   &   2.764   &   2.839   &   2.911   &   3.051   &   3.184   &   3.308   &   3.481   &       &       &   3.123   &       \\
$1 ^1F_3$   &   2.127   &   2.617   &   2.698   &   2.778   &   2.857   &   2.936   &   3.089   &   3.238   &   3.382   &   3.587   &       &       &   3.101   &       \\
\hline \hline
\end{tabular}\\
Relativistic Quark Model (RQM), Blankenbecler- Suger Equation (BSU).
\end{center}
\end{table}
\end{landscape}
\begin{landscape}
\begin{table}[h]
\begin{center}
\caption{Mass spectra (in GeV) of $D_{s}$ meson.}\label{tab:LM04}
\begin{tabular}{lllllllllllllll}
\hline \hline State&\multicolumn{6}{r}{\textbf{
Potential index  $\nu$}}&&&&&Expt.&RQM&RQM&BSU \\
\cline{2-11}
&0.1&0.7&0.8&0.9&1.0&1.1&1.3&1.5&1.7&2.0&\cite{PDG2008}&\cite{Ebert2003}&\cite{Pierro2001}&\cite{Lahde2000} \\
\hline
$1 ^3S_1$   &   2.086   &   2.102   &   2.104   &   2.106   &   2.108   &   2.109   &   2.112   &   2.114   &   2.116   &   2.119   &   2.112   &   2.111   &   2.113   &   2.108   \\
$1^1S_0$    &   2.047   &   1.998   &   1.992   &   1.987   &   1.982   &   1.977   &   1.969   &   1.962   &   1.956   &   1.948   &   1.968   &   1.981   &   1.965   &   1.975   \\
\vspace{-0.1in}\\
$1 ^3P_2$   &   2.165   &   2.348   &   2.372   &   2.394   &   2.416   &   2.436   &   2.474   &   2.506   &   2.535   &   2.573   &   2.572   &   2.560   &   2.581   &   2.586   \\
$1 ^3P_1$   &   2.162   &   2.332   &   2.355   &   2.376   &   2.397   &   2.416   &   2.452   &   2.484   &   2.512   &   2.549   &   2.535   &   2.515   &   2.535   &   2.522   \\
$1 ^3P_0$   &   2.157   &   2.300   &   2.317   &   2.334   &   2.350   &   2.364   &   2.391   &   2.414   &   2.433   &   2.459   &   2.317   &   2.569   &   2.487   &   2.455   \\
$1 ^1P_1$   &   2.163   &   2.337   &   2.360   &   2.382   &   2.402   &   2.422   &   2.457   &   2.488   &   2.516   &   2.552  &   2.460   &   2.508   &   2.605   &   2.502   \\
\vspace{-0.1in}\\
$2 ^3S_1$   &   2.110   &   2.355   &   2.395   &   2.434   &   2.474   &   2.513   &   2.588   &   2.661   &   2.730   &   2.828   &     &   2.716   &   2.806   &   2.722   \\
$2^1S_0$    &   2.101   &   2.303   &   2.334   &   2.365   &   2.395   &   2.425   &   2.482   &   2.535   &   2.585   &   2.654   &       &   2.670   &   2.700   &   2.659   \\
 \vspace{-0.1in}\\
$1 ^3D_3$   &   2.195   &   2.502   &   2.545   &   2.587   &   2.627   &   2.666   &   2.736   &   2.799   &   2.880   &   2.965   &       &       &   2.925   &   2.857   \\
$1 ^3D_2$   &   2.195   &   2.505   &   2.551   &   2.595   &   2.639   &   2.681   &   2.760   &   2.832   &   2.855   &   2.929   &       &       &   2.900   &   2.856   \\
$1 ^3D_1$   &   2.194   &   2.502   &   2.548   &   2.594   &   2.638   &   2.681   &   2.763   &   2.839   &   2.899   &   2.989   &       &       &   2.913   &   2.845   \\
$1 ^1D_2$   &   2.195   &   2.503   &   2.548   &   2.591   &   2.633   &   2.674   &   2.749   &   2.818   &   2.909   &   3.007   &       &       &   2.953   &   2.838   \\
\vspace{-0.1in}\\
$2 ^3P_2$   &   2.136   &   2.489   &   2.549   &   2.608   &   2.668   &   2.727   &   2.843   &   2.954   &   3.040   &   3.190   &       &       &   3.157   &   2.988   \\
$2 ^3P_1$   &   2.134   &   2.478   &   2.535   &   2.593   &   2.651   &   2.708   &   2.820   &   2.928   &   3.062   &   3.217   &       &       &   3.114   &   2.942   \\
$2 ^3P_0$   &   2.132   &   2.455   &   2.507   &   2.560   &   2.612   &   2.663   &   2.763   &   2.858   &   3.032   &   3.182   &       &       &   3.067   &   2.901   \\
$2 ^1P_1$   &   2.135   &   2.482   &   2.540   &   2.598   &   2.656   &   2.713   &   2.826   &   2.935   &   2.949   &   3.078   &       &       &   3.165   &   2.928   \\
\vspace{-0.1in}\\
$3 ^3S_1$   &   2.105   &   2.497   &   2.568   &   2.641   &   2.715   &   2.790   &   2.940   &   3.090   &   3.238   &   3.456   &       &       &   3.345   &   3.087   \\
$3^1S_0$    &   2.101   &   2.459   &   2.523   &   2.587   &   2.652   &   2.717   &   2.846   &   2.973   &   3.096   &   3.275   &       &       &   3.259   &   3.044   \\
 \vspace{-0.1in}\\
$1 ^3F_4$   &   2.214   &   2.630   &   2.693   &   2.755   &   2.814   &   2.872   &   2.979   &   3.075   &   3.221   &   3.361   &       &       &   3.220   &       \\
$1 ^3F_3$   &   2.214   &   2.638   &   2.705   &   2.771   &   2.836   &   2.900   &   3.022   &   3.136   &   3.163   &   3.278   &       &       &   3.224   &       \\
$1 ^3F_2$   &   2.214   &   2.642   &   2.711   &   2.780   &   2.849   &   2.917   &   3.048   &   3.175   &   3.244   &   3.392   &       &       &   3.247   &       \\
$1 ^1F_3$   &   2.214   &   2.635   &   2.702   &   2.766   &   2.830   &   2.892   &   3.010   &   3.119   &   3.295   &   3.466   &       &       &   3.203   &       \\
\hline \hline
\end{tabular}\\
Relativistic Quark Model (RQM), Blankenbecler- Suger Equation (BSU).
\end{center}
\end{table}
\end{landscape}
\section{The Decay constants of the charm flavored mesons} \label{Decay constants}
The decay constants of mesons are important parameters in the study
of leptonic or non-leptonic weak decay processes. The decay
constants of pseudoscalar ($f_P$) and vector ($f_V$) states are
obtained by parameterizing the matrix elements of weak current
between the corresponding mesons and the vacuum as \cite{Quang
Ho-Kim1998}
\begin{table}
\begin{center}
\caption{The decay constants $f_{P/V}$ in MeV of $D$ and $D_s$
systems (The bracketed quantities are with QCD correction).}
\label{tab:LM06}
\begin{tabular}{lrlllllll}
\hline \hline&\m{6}{c} {$D$}&\m{1}{c} {$D_s$}\\
\cline{3-5}\cline{7-9}
\m{0}{c}{}&\m{0}{c}{Models}&\m{0}{c}{1S}&\m{0}{c}{2S}&\m{0}{c}{3S}&&\m{0}{c}{1S}&\m{0}{c}{2S}&\m{0}{c}{3S}\\
\hline $f_P$
    &   CPP$_{\nu}=$0.1     &   154(120)   &   73(57)     &   51(40)   &     &   169(131)   &   80(62)     &   55(43)   \\
    &   0.3                 &   197(155)   &   111(88)    &   85(67)   &     &   216(167)   &   122(95)    &   93(72)   \\
    &   0.5                 &   227(178)   &   141(111)   &   113(89)   &    &   249(193)   &   155(120)   &   124(96)   \\
    &   0.7                 &   248(195)   &   166(131)   &   137(108)   &   &   273(211)   &   184(142)   &   152(118)   \\
    &   0.8                 &   257(202)   &   177(139)   &   148(117)   &   &   283(219 )  &   196(152)   &   165(128)   \\
    &   0.9                 &   265(208)   &   188(148)   &   159(125)   &   &   291(226)   &   208(161)   &   177(137)   \\
    &   1.0                 &   272(213)   &   197(155)   &   169(133)   &   &   299(232)   &   219(170)   &   189(146)   \\
    &   1.1                 &   278(218)   &   207(162)   &   178(140)   &   &   306(237)   &   230(178)   &   200(155)   \\
    &   1.3                 &   289(226)   &   223(175)   &   196(154)   &   &   318(246)   &   249(193)   &   221(171)   \\
    &   1.5                 &   297(233)   &   238(187)   &   212(167)   &   &   327(254)   &   267(207)   &   240(186)   \\
    &   RQM\cite{Ebert2006}   &   234     &           &           &   &   268     &           &           \\
    &   BS\cite{Cvetic2004,Wang2006}  &   230$\pm$25     &           &           &   &   248$\pm$27       &           &           \\
\cline{3-5}\cline{7-9}
   &\m{6}{c} {$D^{*}$}&\m{1}{c} {$D_s^{*}$}\\
\cline{3-5}\cline{7-9}
$f_V$   &   CPP$_{\nu}=$0.1 &  156(104)   &  73(49)     &   51(34)      &   &   170(116)   &   80(54)     &   55(38)   \\
    &                   0.3 &  202(135)   &  112(75)    &   85(57)      &   &   219(149)   &   123(84)    &   93(63)   \\
    &                   0.5 &  234(157)   &  143(96)    &   114(76)     &   &   254(173)   &   157(107)   &   125(85)   \\
    &                   0.7 &  258(173)   &  169(113)   &   139(93)     &   &   280(190)   &   186(126)   &   153(104)   \\
    &                   0.8 &  268(180)   &  181(121)   &   150(101)    &   &   290(198)   &   199(135)   &   166(113)   \\
    &                   0.9 &  277(186)   &  192(128)   &   161(108)    &   &   300(204)   &   211(144)   &   179(122)   \\
    &                   1.0 &  285(191)   &  202(135)   &   172(115)    &   &   308(210)   &   223(152)   &   191(130)   \\
    &                   1.1 &  292(196)   &  212(142)   &   182(122)    &   &   316(215)   &   234(159)   &   203(138)   \\
    &                   1.3 &  304(204)   &  230(154)   &   201(134)    &   &   329(224)   &   255(173)   &   224(153)   \\
    &                   1.5 &  314(211)   &  247(165)   &   218(146)    &   &   340(231)   &   273(186)   &   244(166)   \\
    &   RQM\cite{Ebert2006}   &   310               &&&&                      315                             \\
    &   BS\cite{Cvetic2004,Wang2006}  &   340$\pm$23     &           &           &   &   375$\pm$24       &           &           \\
    \hline \hline
\end{tabular}
\end{center}
\end{table}
\begin{equation}
 \langle 0 | \bar q \gamma^\mu \gamma_5 c |
P_{\mu}(k)\rangle = i f_P k^\mu
\end{equation}
\begin{equation}
\langle 0 | \bar q \gamma^\mu c | V(k,\epsilon)\rangle = f_V M_V
\epsilon^\mu
\end{equation}
where $k$ is the meson momentum, $\epsilon^\mu$ and $M_V$ are the
polarization vector and mass of the vector meson.\\ In the
relativistic quark model, the decay constant can be expressed
through the meson wave function $\Phi_{P/V}(p)$ in the momentum
space \cite{Ebert2003}.
\begin{table}
  \centering
  \caption{Psedoscalar $f_P$ decay constants for  $D$  mesons in (MeV)}\label{tab:LM07}
  \begin{tabular}{rlllllllllll}
\hline \hline
  &\m{2}{c}{$f_{P}(D)$}&&\m{2}{c}{$f_{P}(D_s)$}&&\m{2}{c}{$f_{P}(D_s)$/$f_{P}(D)$}\\
  \cline{2-3}\cline{5-6} \cline{8-9}
&\m{0}{c}{Our}&\m{0}{c}{Others}&&\m{0}{c}{Our}&\m{0}{c}{Others}&&\m{0}{c}{Our}&\m{0}{c}{Others} \\
     \hline
CPP$_{\nu}$=0.1 &   154(120)   &                                             &&  169(131)   &                                           &&   1.097(1.084)   &       \\
0.3 &   197(155)   &   230\cite{Gvetic2004}                                  &&   216(167)   &   248\cite{Gvetic2004}                   &&  1.096(1.082)   &   1.08\cite{Gvetic2004}    \\
0.5 &   227(178)   &   234\cite{Ebert2006}                                   &&   249(193)   &   268\cite{Ebert2006}                    &&   1.096(1.083)   &   1.15 \cite{Ebert2006}       \\
0.7 &   248(195)   &   203\cite{Narison2001}                                 &&   273(211)   &   235\cite{Narison2001}                  &&   1.098(1.084)   &   1.15\cite{Narison2001}   \\
0.8 &   257(202)   &   208\cite{Follana2007}                                 &&   283(219)   &   241\cite{Follana2007}                  &&   1.098(1.084)   &   1.164\cite{Follana2007}   \\
0.9 &   265(208)   &   201\cite{Aubin2005}                                   &&   291(226)   &   249\cite{Aubin2005}                    &&   1.098(1.085)   &   1.24\cite{Aubin2005}  \\
1.0 &   272(213)   &   206\cite{Khan2007}                                    &&   299(232)   &   220\cite{Khan2007}                     &&  1.099(1.086)   &   1.07\cite{Khan2007}   \\
1.1 &   278(218)   &   235\cite{Chiu2005}                                    &&  306(237)   &    266\cite{Chiu2005}                     &&  1.099(1.086)   &   1.13\cite{Chiu2005}   \\
1.3 &   289(226)   &   223\cite{Rosner2008}                                  &&   318(246)   &   276\cite{Rosner2008}                   &&  1.101(1.087)   &   1.23\cite{Rosner2008}   \\
1.5 &   297(233)   &                                                         &&   327(254)   &                                          &&  1.101(1.088)   &       \\
\hline
   \hline
   \end{tabular}
\end{table}
\begin{eqnarray} f_{P/V} & = &
\dis\sqrt{\frac{12}{M_{P/V}}}\int\dis\frac{d^3p}{(2\pi)^3} \left (
\frac{E_c(p) + m_c}{2E_c(p)}\right )^{1/2} \left ( \frac{E_{\bar
q}(p) + m_{\bar q}}{2E_{\bar q}(p)}\right )^{1/2} \cr && \left \{1+
\lambda_{P/V} \frac{p^2}{[E_c(p)+m_c][E_{\bar q}(p)+m_{\bar
q}]}\right \}\Phi_{P/V}(p) \label{tab:pseudj}\end{eqnarray} with
$\lambda_P=-1$ and $\lambda_V=-1/3$. In the nonrelativistic limit
$\dis\frac{p^2}{m^2}<<1.0$, this expression reduces to the well
known relation between  $f_{P/V}$ and the ground state wave function
at the origin $R_{P/V}(0)$ the Van-Royen-Weisskopf formula
\cite{Vanroyenaweissskopf}. Though most of the models predict the
meson mass spectrum successfully, there exist wide range of
predictions of their  decay constants. For example, the ratio
$\frac{f_P}{f_V}$ was predicted to be $>1$ in most of the
nonrelativistic cases, as $m_P<m_V$ and their wave function at the
origin has assumed to be  as $R_P(0) \sim R_V(0)$ \cite{Hwang1997}.
The ratio computed in the relativistic models \cite{Wang2006} have
predicted $\frac{f_P}{f_V}<1$, particularly in the $Q \bar Q$
sector, but $\frac{f_P}{f_V}>1$ in the heavy-light flavour sector.
The disparity of the predictions of these decay constants play
decisive role in the precision measurements of the weak decay
parameters. The value of the radial wave function ($R_{P}$) for
$0^{-+}$ and ($R_{V}$) for $1^{--}$ states would be different due to
their spin dependent hyperfine interaction. The spin hyperfine
interaction of the heavy flavour mesons are small and this can cause
a small shift in the value of the wave function at the origin.
Though, many models neglect this difference between $(R_{P})$ and
$(R_{V})$, we consider this correction by making an ansatz that the
$R_{P/V}(0)$  are related to the value of the radial wave function
at the origin, $R_{n}(0)$ according to the same
way their masses are related. Thus, by considering 
\begin{equation}
M_{nP/V}=M_{n,CW}\left[1+(SF)_{P/V} \frac{\langle
V_{SS}\rangle_{n}}{M_{n,CW}}\right]
\end{equation}
and following the fact that any $c$-number, $a$, commutes with the
Hamiltonian, \emph{i.e.} $aH\Psi=H(a\Psi)$, we express,
 \begin{equation}\label{Rcw1}
R_{nP/V}(0)=R_{n}(0)\left[1+(SF)_{P/V}
\frac{(M_{nV}-M_{nP})}{M_{n,CW}} \right]
\end{equation}
Here $(SF)_P=-\frac{3}{4}$ and $(SF)_V=\frac{1}{4}$ are the spin
factor corresponding to the pseudoscalar ($J=0$) spin coupling and
vector ($J=1$) spin coupling respectively \cite{Bhavin2009}.
$M_{n,CW}$ and $R_{n}(0)$ are spin average mass and the normalized
spin independent wave function at the origin of the meson state
respectively. It can easily be seen that this expression given by
Eqn \ref{Rcw1} is consistent with the relation
\begin{equation}
R(0)=\frac{3 R_{V}(0)+R_{P}(0)}{4}
\end{equation}
given by \cite{AjayRai2008,Bodwin1995} for  $nS$ states.
 The decay
constants by incorporating first order QCD correction to the Van
Royen-Weiskopff formula are given by
\cite{Berezhnoy1996,EBraaten1995},
\begin{equation}
f^2_{P/V}(nS)=\frac{3 \left| R_{nP/V}(0)\right|^2} { \pi M_{nP/V}}
{\bar C^2}(\alpha_s) \label{eq:fpv}
 \end{equation}
where, the first order QCD correction factor, $\bar{C}(\alpha_{s}$)
is expressed for the $Q \bar{q}$ system as
\begin{equation}
{\bar C}(\alpha_s)=1+\frac{\alpha_s}{\pi}
\left[\frac{m_1-m_{2}}{m_1+m_{2}} {\rm \
 ln} \frac{m_1}{m_{2}}-
\delta^{V,P} \right] \label{eq:c2}
\end{equation}
Here $\delta^{V} = \frac{8}{3}  $ \cite{Berezhnoy1996,Gerstein1998}
and $\delta^P=2$ \cite{Berezhnoy1996,EBraaten1995,Gerstein1998}. For
the $Q \bar q$ system, $m_1=m_Q$ and $m_2=m_{\bar q}$. We re-examine
the predictions of the decay constants $f_P$ and $f_V$ under
different potential schemes (by the choices of different $\nu$) with
and without the QCD correction. Our computed results up to $3S$
states of the $D$ and $D_s$ systems are tabulated in Tables
\ref{tab:LM06}. The ratio of $\frac{f_P(D_s)}{f_P(D)}$ for $1S$
state is tabulated against different choices of $\nu$ in Table
\ref{tab:LM07}. The present results are in accordance with other
predictions as seen from the the pseudoscalar decay constant $f_{D}$
and $f_{D_s}$.
\begin{table}
\begin{center}
\caption{The root mean square radii (in $fm$)  of $D$ and $D_s$
systems.} \label{tab:LM13}
\begin{tabular}{crlllllll}
\hline \hline
\m{0}{c}{State}&\m{0}{c}{Model}&\m{0}{c}{1S}&\m{0}{c}{2S}&\m{0}{c}{3S}&\m{0}{c}{1P}&\m{0}{c}{2P}&\m{0}{c}{1D}&\m{0}{c}{1F}\\
  \hline
  &CPP$_\nu=$    0.1 &   1.344   &   3.991   &   7.202   &   3.000   &   6.124   &   4.647   &   6.262   \\
$D$&   0.3 &   1.043   &   2.687   &   4.429   &   1.966   &   3.702   &   2.799   &   3.587   \\
&   0.5 &   0.903   &   2.136   &   3.347   &   1.567   &   2.800   &   2.140   &   2.669   \\
&   0.7 &   0.818   &   1.813   &   2.734   &   1.343   &   2.300   &   1.781   &   2.178   \\
&   0.8 &   0.787   &   1.696   &   2.515   &   1.263   &   2.124   &   1.656   &   2.008   \\
&   0.9 &   0.761   &   1.597   &   2.334   &   1.198   &   1.978   &   1.553   &   1.869   \\
&   1.0 &   0.738   &   1.514   &   2.182   &   1.143   &   1.856   &   1.467   &   1.754   \\
&   1.1 &   0.719   &   1.442   &   2.051   &   1.095   &   1.752   &   1.394   &   1.657   \\
&   1.3 &   0.687   &   1.323   &   1.840   &   1.019   &   1.584   &   1.278   &   1.502   \\
&   1.5 &   0.663   &   1.231   &   1.678   &   0.961   &   1.455   &   1.189   &   1.385   \\
\vspace{-0.1in}\\
&CPP$_\nu=$   0.1 &   1.255   &   3.749   &   6.777   &   2.819   &   5.780   &   4.381   &   5.914   \\
$D_s$&   0.3 &   0.979   &   2.532   &   4.182   &   1.855   &   3.500   &   2.646   &   3.395   \\
&   0.5 &   0.848   &   2.015   &   3.161   &   1.479   &   2.646   &   2.023   &   2.524   \\
&   0.7 &   0.769   &   1.710   &   2.581   &   1.267   &   2.172   &   1.682   &   2.058   \\
&   0.8 &   0.740   &   1.599   &   2.374   &   1.192   &   2.005   &   1.564   &   1.897   \\
&   0.9 &   0.715   &   1.506   &   2.203   &   1.130   &   1.867   &   1.466   &   1.766   \\
&   1.0 &   0.694   &   1.427   &   2.058   &   1.077   &   1.751   &   1.385   &   1.656   \\
&   1.1 &   0.676   &   1.359   &   1.935   &   1.033   &   1.652   &   1.315   &   1.564   \\
&   1.3 &   0.646   &   1.247   &   1.735   &   0.960   &   1.493   &   1.205   &   1.417   \\
&   1.5 &   0.623   &   1.159   &   1.581   &   0.905   &   1.371   &   1.120   &   1.305   \\
 \hline \hline
\end{tabular}
\end{center}
\end{table}
\begin{table}
\begin{center}
\caption{Mean square velocity of the quark within $D$ and $D_s$
states.} \label{tab:LM11}
\begin{tabular}{crlllllll}
\hline \hline
\m{0}{c}{State}&\m{0}{c}{Model}&\m{0}{c}{1S}&\m{0}{c}{2S}&\m{0}{c}{3S}&\m{0}{c}{1P}&\m{0}{c}{2P}&\m{0}{c}{1D}&\m{0}{c}{1F}\\
  \hline
    &CPP$_\nu=$     0.1 &   0.197   &   0.091   &   0.066   &   0.098   &   0.069   &   0.077   &   0.069   \\
$D$&   0.3 &   0.307   &   0.208   &   0.184   &   0.220   &   0.190   &   0.208   &   0.208   \\
&   0.5 &   0.397   &   0.334   &   0.331   &   0.341   &   0.332   &   0.353   &   0.372   \\
&   0.7 &   0.472   &   0.469   &   0.504   &   0.459   &   0.492   &   0.506   &   0.556   \\
&   0.8 &   0.506   &   0.539   &   0.599   &   0.517   &   0.578   &   0.584   &   0.653   \\
&   0.9 &   0.537   &   0.610   &   0.700   &   0.573   &   0.666   &   0.662   &   0.753   \\
&   1.0 &   0.566   &   0.681   &   0.805   &   0.628   &   0.757   &   0.740   &   0.853   \\
&   1.1 &   0.593   &   0.754   &   0.914   &   0.681   &   0.850   &   0.818   &   0.955   \\
&   1.3 &   0.642   &   0.899   &   1.142   &   0.783   &   1.041   &   0.971   &   1.160   \\
&   1.5 &   0.685   &   1.043   &   1.381   &   0.878   &   1.234   &   1.120   &   1.363   \\
\vspace{-0.1in}\\
&CPP$_\nu=$    0.1 &   0.133   &   0.060   &   0.043   &   0.065   &   0.046   &   0.051   &   0.045   \\
$D_s$&   0.3 &   0.205   &   0.136   &   0.121   &   0.145   &   0.124   &   0.136   &   0.135   \\
&   0.5 &   0.263   &   0.219   &   0.217   &   0.224   &   0.217   &   0.231   &   0.243   \\
&   0.7 &   0.313   &   0.308   &   0.330   &   0.302   &   0.323   &   0.331   &   0.364   \\
&   0.8 &   0.335   &   0.354   &   0.393   &   0.340   &   0.379   &   0.383   &   0.428   \\
&   0.9 &   0.355   &   0.401   &   0.459   &   0.377   &   0.437   &   0.434   &   0.493   \\
&   1.0 &   0.375   &   0.448   &   0.528   &   0.413   &   0.497   &   0.486   &   0.560   \\
&   1.1 &   0.393   &   0.496   &   0.600   &   0.448   &   0.559   &   0.537   &   0.627   \\
&   1.3 &   0.425   &   0.592   &   0.751   &   0.516   &   0.684   &   0.639   &   0.762   \\
&   1.5 &   0.454   &   0.687   &   0.909   &   0.579   &   0.812   &   0.737   &   0.897   \\
\hline \hline
\end{tabular}
\end{center}
\end{table}
\section{Root mean square Radii of the $D$ and $D_s$ meson states and Average quark Velocities}
The mean square size of the meson is an important parameter in the
estimations of hadronic transition widths
\cite{Gottfried1978,Voloshin1979,Kuang1990}, while the average
velocity of the quarks within a quark-antiquark bound state is
important for the estimation of the relativistic corrections and
useful particularly in the NRQCD formalism. It is also important in
the estimation of their production rates \cite{Bodwin2008}. Thus
with our numerical radial wave functions obtained for different
choices of the potential index $\nu$, we compute the mean square
radii of the  meson state as
 \begin{equation}
\left\langle r^2\right\rangle_{nl}= \int_{0}^{\infty} r^4
|R_{nl}(r)|^2 dr \label{eq:meansqradii}
\end{equation}
and the average mean square  velocity of the quark$/$antiquark
inside the state as \cite{Juan-Luis2008}
\begin{equation}
 \left \langle  {\frac{{v}^2}{c^2}} \right\rangle_{nl}= \frac{1}{2 M_1}(E_{nl}-\langle
V(r)\rangle_{nl})\end{equation} Here,
$E_{nl}$ is the binding energy of the $n \ell^{th}$ state and
$\langle V(r)\rangle_{nl}$ is the expectation value of the
quark-antiquark interaction (without spin dependent terms)
potential energy in that state. The computed root mean square radii
and the relative mean square velocities of the bound states within
the mesons are tabulated
 in Table \ref{tab:LM13} and Table \ref{tab:LM11}  with different
 choices of $\nu$ respectively.
\section{Inclusive Semileptonic Decay of Open Charm Flavour
Mesons} 
\begin{figure}[h]
\begin{center}
\includegraphics[width=8cm]{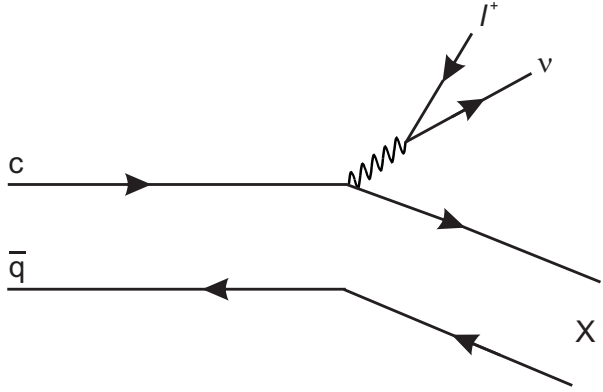}\\
\caption{Feynman diagram for semileptonic $D$ Decay } \label{Dsfig}
\end{center}
\end{figure}

Inclusive widths of the heavy flavor hadrons are examples of
the genuine short-distance processes. The open charm mesons, decay
through,  $c\to q\ell^+\nu$, where $q=d,~s$. The light $d$ or $s$
daughter quark is bound to the initial light quark of the charm
meson by the strong interaction to form a new hadron $X$, according
to the Feynman diagram of Fig.\ref{Dsfig}.\\
In semileptonic decays, the two leptons do not feel the strong
interaction, and are thus free of strong binding effects. Therefore,
they can be factored out of the hadronic matrix element in the
amplitude of the semileptonic decay process as
\begin{equation}
A=\frac{G_F}{\sqrt{2}}V_{cq}^*
\bar{\nu}\gamma_\mu(1-\gamma_5)l\langle X|\bar{q}\gamma^\mu
(1-\gamma_5)c|D\rangle ,
\end{equation}
where all strong interactions are included in the hadronic matrix
element $\langle X|\bar{q}\gamma^\mu (1-\gamma_5)c|D\rangle $. The
amplitude of the semileptonic decay process depends both on the
hadronic matrix element and the quark-mixing parameter $V_{cq}$--the
Cabibbo-Kobayashi-Maskawa (CKM) matrix element. Thus, the
semileptonic charm meson decay process is a good laboratory for both
studying the quark-mixing mechanism and testing theoretical
techniques developed for calculating the hadronic matrix element.
The hadronic matrix element can be decomposed into several form
factors according to its Lorentz structure. The form factors are
generally controlled by non-perturbative dynamics,  since
perturbative QCD could not be applied directly.
\\For the present study, we compute the decay width of the $D$ and
$D_s$ mesons using the nonperturbative bound state effects. The
decay width and branching ratio
 for the $\Gamma( D \rightarrow \bar K^0+ e^+ +\nu_e)$  and  $\Gamma ( D_s \rightarrow
\phi+ \ell^+ + \nu_\ell)$ mesons are calculated using the expression
given by  \cite{Yosef1989,Cabibbo1978,Luke1993},
 \be \label{eq:qcdcorrDd}
\Gamma_{sl}(D)= \frac{G^2_F m_c^5}{192 \pi^3}
(|V_{cs}|^2+|V_{cd}|^2) \ \left[f(x)-\frac{\alpha_s}{\pi}
g(x)\right]\ee
 \be \label{eq:qcdcorrDs}
\Gamma_{sl}(D_s)= \frac{G^2_F m_c^5}{192 \pi^3} |V_{cs}|^2 \
\left[f(x)-\frac{\alpha_s}{\pi} g(x)\right]\ee where $f(x)=1-8x+8
x^3-x^4-12x^2\ \log x$, and the analytic expression of the function
$g(x)$ is given by  \cite{Quang Ho-Kim1998,Yosef1989}
\begin{equation}\label{g}
  g(x)= -{15.28} x^6+{48.68} x^5-{60.06} x^4+{35.3}
   x^3-{8.11} x^2-{1.97} x+{2.41}
\end{equation}
 Here, the parameter
$x$ is computed as  $x=m_{s}^{2}/(m_c^{eff})^2$. Generally, for the
calculation of the semileptonic decay of the heavy flavour mesons,
the  $m_s$ is taken as the model mass parameters coming from the
fitting  of its mass spectrum. However, taking into account of the
binding energy effects of the decaying heavy quark within the
potential confinement scheme, we consider the decaying heavy quark
mass as the effective mass of the quarks, $m_{q}^{eff}$.
Accordingly, we define the effective masses of the quarks in the $Q
\bar q$ system as
\begin{eqnarray}\label{eq:417}
\nonumber m^{eff}_{Q}&=&m_Q\left(
1+\frac{\left<E_{bind}\right>}{m_{Q}+ m_{\bar{q}}}\right) \\
m^{eff}_{\bar q}&=&m_{\bar q} \left(
1+\frac{\left<E_{bind}\right>}{m_{Q}+ m_{\bar{q}}}\right)
\end{eqnarray}
to account for its bound state effects. The binding effect has been
calculated as $\left<E_{bind}\right>=M_{Q\bar{q}}-(m_{Q}+m_{\bar
q})$,  where $m_Q$ and $m_{\bar q}$ are the model mass parameters
employed in its spectroscopic study and $M_{Q \bar {q}}$ is the mass
of the mesonic state. The effective mass of the quarks would be
different from the adhoc choices of the model mass parameters. For
example, within the meson the mass of the quarks may get modified
due to its binding interactions with other quark. Thus, the
effective mass of the  charm quark will be different when it is in
$c\bar{s}$ combinations or in $c\bar{d}$ combinations due to the
residual strong interaction effects of the bound
systems.\\
\begin{table} \caption{The inclusive semileptonic BR of $ D \rightarrow \bar K^0+ e^+ +\nu_e$ and $D_s
\rightarrow \phi+ \ell^+ + \nu_\ell$ states in \%.} \label{tab:LM09}
\begin{center}
\begin{tabular}{cccccccccccc}
  \hline \hline
CPP$_\nu\rightarrow$  & 0.1   &   0.3  &   0.5     &   0.7    &   0.8  &   0.9     &   1.0     &   1.1     &   1.3     &   1.5&Expt.\cite{PDG2008}        \\
\hline
BR$_{D}$    &7.4    &   6.7  &   6.1     &   5.6    &   5.4  &   5.3     &   5.1     &   5.0     &   4.7     &   4.5&  8.6$\pm$0.5       \\
BR$_{D_{s}}$&2.77   &   2.57 &   2.42    &   2.29   &   2.24 &   2.19    &   2.15    &   2.11    &   2.04    &   1.98&2.36$\pm$0.26        \\
  \hline\hline
\end{tabular}
\end{center}
\end{table}
From the computed inclusive semileptonic decay widths, the Branching
ratio of $D_q$ mesons are taken here from the relation
\begin{equation} \label{eq:BR}
BR= \Gamma_{sl}\times \tau,
\end{equation} The Lifetime of these mesons ($\tau_{D}=1.04$ $ps^{-1}$ and
$\tau_{D_s}=0.5$ $ps^{-1}$) are obtained from the world average
value reported by Particle Data Group (PDG-2008)\cite{PDG2008}. The
computed results of $D$ and $D_s$ mesons are listed in Table:
\ref{tab:LM09}. Our results are found to be in agreement with
experimental results at lower potential indexes $\nu\approx0.1$ to 0.5, which
in consistent with the agreement observed for their spectroscopy.
\section{Leptonic Decay of the Open Heavy Flavour
Mesons}\label{Lep} 
\begin{figure}[h]
 \begin{center}
\includegraphics{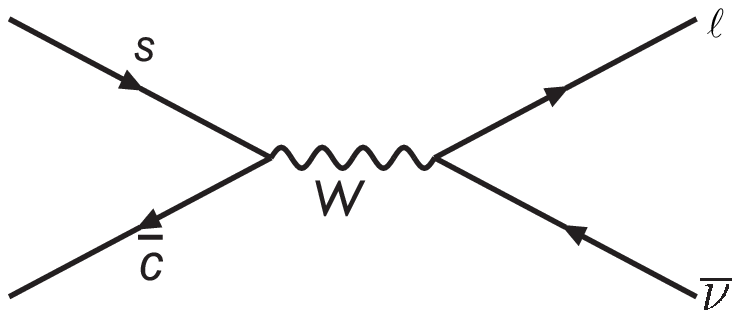}
  \end{center}
 \caption{Feynman diagram in standard model for
 $D_{s}$ ${\to}$ $\ell$ $\overline{\nu}$ decay.}\label{Fig:rad1}
 \end{figure}

Charged mesons formed from a quark and anti-quark
can decay to a charged lepton pair when these objects annihilate via
a virtual $W^{\pm}$ boson (See Fig.\ref{Fig:rad1}).
 \begin{table}[b]
\begin{center}
\caption{The leptonic BR of $D$ and $D_s$  mesons.}
\label{tab:LM10}
\begin{tabular}{rrccc}
\hline \hline
&&\m{0}{c}{BR$_{\tau}\times10^{-3}$}&\m{0}{c}{BR$_{\mu}\times10^{-4}$}
&\m{0}{c}{BR$_{e}\times10^{-8}$}\\
  \hline
$D$&CPP$_\nu=$0.1 &   1.5 &       2.2 &        0.5            \\
               &0.3 &   1.7 &       3.6 &        0.8                   \\
               &0.5 &   1.6 &       4.7 &        1.1                    \\
               &0.7 &   1.3 &       5.6 &        1.3                   \\
               &0.8 &   1.2 &       5.9 &        1.4                   \\
               &0.9 &   1.0 &       6.3 &        1.5                  \\
               &1.0 &   0.9 &       6.6 &        1.5                    \\
               &1.1 &   0.7 &       6.9 &        1.6                   \\
               &1.3 &   0.5 &       7.3 &        1.7                    \\
               &1.5 &   0.3 &       7.7 &        1.8                 \\
&Expt.\cite{PDG2008}&$<$ 2.1&4.4$\pm$0.7& \\

 \hline &&\m{0}{c}{BR$_{\tau}\times10^{-2}$}&\m{0}{c}{BR$_{\mu}\times10^{-3}$}
 &\m{0}{c}{BR$_{e}\times10^{-7}$}\\
  \hline
$D_s$&CPP$_\nu=$0.1 &   4.3 &       2.5 &       0.6              \\
               &0.3 &   6.3 &       4.1 &       1.0                    \\
               &0.5 &   7.4 &       5.4 &       1.3                    \\
               &0.7 &   8.0 &       6.4 &       1.5                    \\
               &0.8 &   8.2 &       6.9 &       1.6                    \\
               &0.9 &   8.3 &       7.3 &       1.7                    \\
               &1.0 &   8.4 &       7.7 &       1.8                    \\
               &1.1 &   8.4 &       8.0 &       1.9                 \\
               &1.3 &   8.4 &       8.6 &       2.0                  \\
               &1.5 &   8.3 &       9.1 &       2.1                   \\
&Expt.\cite{PDG2008}&6.6$\pm$0.6&6.2$\pm$0.6&\\

\hline
 \hline
\end{tabular}
\end{center}
\end{table}
quark-antiquark annihilations via a virtual $W^+(W^-)$ to the
$\ell^+ \nu(\ell^-\nu)$ final states occur for the
$\pi^{\pm},K^{\pm},D_{s}^{\pm}$ and $B^{\pm}$ mesons. There are
several reasons for studying the purely leptonic decays of charged
mesons \cite{Rosner2008}. Such processes are rare but they have
clear experimental signatures due to the presence of a highly
energetic lepton in the final state. The theoretical predications
are very clean due to the absence of hadrons in the final state
\cite{Stefano2007}.
The total leptonic width of $D, D_s$ mesons are given by \beq
\Gamma(D_q^+\to \ell^+ \nu_l) = \frac{G_F^2}{8\pi} f^2_{D_q}
|V_{cq}|^2 m_\ell^2 \left(1-\frac{m_\ell^2}{M_{D_q}^2}\right)^2
M_{D_q}, \ \ q = d, s \label{DQLNU} \eeq
 These transitions are helicity
suppressed \index{helicity suppression}; \emph{i.e.}, the amplitude
is proportional to $m_\ell$, the mass of the lepton $\ell$, in
complete analogy  to $\pi^+ \to \ell^+\nu$.\\
The leptonic widths of the charged $D$ and $D_s$ ( $1 ^1S_{0}$
state) mesons are obtained using Eqn.\ref{DQLNU} employing the
predicted values of the pseudoscalar decay constants $f_{D}$ and
$f_{D_s}$ along with the masses of the $M_{D}$ and $M_{D_s}$
obtained from the CPP$_\nu$ model. The leptonic widths for separate
lepton channel by the choice of $m_{\ell=\tau,\mu,e}$ are computed.
The branching ratios of the total leptonic widths are then
calculated using Eqn. \ref{eq:BR}. The present results as tabulated
in Table \ref{tab:LM10} are in accordance with the available
experimental values.
\section{Result and Discussions}
The spectroscopic results obtained for open charm ($D, D_s$) mesons
with different choice of the confining potential index $\nu$ from
0.1 to 2.0 are tabulated along with other relativistic quark model
predictions and with the known experimental states. Our predicted
masses of $P-$wave $D-$meson state $1 ^3P_2$(2342 - 2514 MeV), $1
^3P_1$(2361 - 2542 MeV), $1 ^3P_0$(2335 - 2510 MeV) and $1
^1P_1$(2269 - 2385 MeV) for the choices of $\nu$,
$1.0\leq\nu\leq2.0$ are in a accordance with other theoretical model
prediction \cite{Pierro2001,Ebert2003,Lahde2000}. Similar agrement
for the predicted masses for the $2S$, $1D$, $2P$, $3S$ and $1F$
states are also observed in the same range of $\nu$ values. While
the experimental candidate for J$^{P}=2^{+}$ $D^{*}_{1}$(2460),
J$^{P}=1^{+}$ $D$(2420) \cite{PDG2008} and
  J$^P=0^{+}$ state observed in the range 2300 - 2400 MeV
   by Belle and Focus \cite{Link:2003bd} lie within our predicted range.
In the case of open strange-charm mesons ($D_s$), our predictions
within the range of $\nu$, $1.5\leq\nu\leq2.0$ for the $1P$, $2S$,
$1D$, $2P$, $3S$ and $1F$ states of the $D_s$ mesons are in
accordance with other theoretical model prediction
\cite{Pierro2001,Ebert2003,Lahde2000}. In particular the
experimental states of $D_{s2}(2573)$ \cite{PDG2008} lie within our
predicted range of $1 ^3P_2$ (2416-2573), $D_{s1}$ (2460)
\cite{PDG2008} lies in the predicted range of $1 ^3P_1$ (2397-2549
MeV) for $1.0\leq\nu\leq2.0$ and the $D^{*}_{s0}$ (2317)
\cite{PDG2008} lies close to the predicted range of $1 ^3P_0$
(2317-2350 MeV) for $0.8\leq\nu\leq1.0$ of the CPP$_\nu$ model. The
radial excitation of $D_s^{*}$(2715) by the Belle
group\cite{Abe:2006xm} is found to be close to the predicted  $3
^3S_1$ state for $\nu=1.0$.  Even higher excited states  of $c \bar
s$ system has been observed by the BaBar collaboration
\cite{Aubert:2006mh} with spin parity $0^{+},1^{+}$ and $2^{+}$
\emph{etc.,} with mass at 2856$\pm$1.5$\pm$5.0 which in our case
corresponds to the $2P$ state with the predicted mass range of $2
^3P_2$(2668 - 2954 MeV),  $2 ^3P_1$(2651 - 2928 MeV), $2 ^3P_0$(2612
- 2858 MeV)  and  $2 ^1P_1$(2656 - 2935 MeV) with the choices of
$\nu$ in the range $1.0\leq\nu\leq1.5$. Thus the present study on
the open charm and open strange-charm mesons using CPP$_\nu$ model,
we have been able to identify the recently discovered $D-$meson
states as well as the $D_s-$meson states. Other predicted high
angular momentum states $\ell\geq2$  of these mesons are expected to
be seen in the future experiments at BES-III, BaBar, Belle and CLEO
collaborations. Our $1F-$ state mass predictions are in accordance
with the theoretical predictions based on a relative quark model
\cite{Pierro2001}
but at higher choice of $\nu$ ($\nu\geq1.5$).  \\
Our results for $f_P$ and $f_V$  in the potential index ranging from
0.5 to 1.5 are fairly close to the known theoretical prediction as
seen from Tables \ref{tab:LM06}. The present tabulated results with
QCD corrections (shown in brackets) are in agreement with the
experimental values but higher potential index beyond $\nu=1.0$.
CLEO has reported the first significant measurement of
$f_{D^{+}}=222.6\pm16.7$ MeV \cite{Artuso:2005ym} which is close to
our predicted value of 227 MeV (without QCD corrections)for $\nu =$
0.5 and that 226 MeV (with QCD corrections) obtained at $\nu=1.3$.
The accuracy of the previous world average has been improved by
BaBar with $f_{Ds}=283\pm17$ MeV \cite{Aubin:2005ar} which within
the range of values 273 - 283 MeV predicted here for the potential
index $0.7\leq\nu\leq0.8$ without QCD correction and goes beyond
$\nu=1.5$ with QCD correction . However the ratio
$f_{P}(D_s(1S))/f_{P}(D(1S))$ is very close to each other between
1.09 to 1.10 (without the QCD correction) and between 1.082-1.088
(with QCD correction) with changing $\nu$  from 0.1 to 1.5. The
ratio predicted by the CPP$_\nu$ model is thus very close to the
ratio predicted by \cite{Gvetic2004} and \cite{Khan2007}
but is lower than the ratio of 1.27 as per the recent experimental values of CLEO \cite{Artuso:2005ym} and BaBar \cite{Aubin:2005ar}.\\
The semileptonic branching ratios of $D$ and $D_s$ mesons computed
here using CPP$_\nu$ model (See Table \ref{tab:LM09}) are all found
to be in good agreement with their respective experimental results.
It can also be seen that the results do not vary appreciably with
change in potential index $\nu$, indicating lesser influence of
strong interaction effects in these decays. Though our predictions
for  $D_s$ are well within the experimental error bar, the
branching ratio of $D-$meson is slightly under estimated.\\
Present study on the leptonic decay branching ratios of $D$ and $
D_s$ system  presented in Table \ref{tab:LM10} are as per the
available experimental limits. The branching ratio in $\tau-$lepton
channel for $D$ and $D_s$ mesons lie within the predicted range  for
the potential index $\nu\approx$ 0.3 to 0.5. In the case of
$\mu-$lepton channel, the experimental value of
($4.4\pm0.7)\times10^{-4}$ for $D-$meson lie in the predicted range
for the potential index $\nu$ = 0.3 to 0.5 and that for $D_s$ meson
in the potential index $\nu$ = 0.7 to 0.8. Large experimental
uncertainty in the electron channel make it difficult for any
reasonable conclusion. Probably, future high luminosity better
statistics and high confidence level data sets will be able to
provide more
light on the spectroscopy and decay properties of these open charm mesons.\\
\\
\textbf{Acknowledgement:} Part of this work is done with a financial
support from DST, Government of India, under a Major Research
Project \textbf{SR/S2/HEP-20/2006}.\\


\begin{thebibliography}{99}
\bibitem{Godfrey1991}  Godfrey S  and  Kokoski R, Phys. Rev. {\bf D43}, 1679 (1991).
\bibitem{Pierro2001} M. Di Pierro \emph{et al.}, Phys. Rev. {\bf D64}, 114004
(2001).
\bibitem{ebert1998}  Ebert D, Faustov R N and  Galkin V O, Phys. Rev. {\bf D57}, 014027 (1998).
\bibitem{Bardeen2003}  Bardeen W A, Eichen E J and Hill C T, Phys. Rev. {\bf D68}, 054024 (2003).
\bibitem{Colangelo2003} Colangelo P,  Fazio F De and Ferrandes R, Nucl. Phys. (Proc. Suppl.){\bf D163}, 177 (2007).
\bibitem{Falk2003} Falk A F and Mehen T, Phys. Rev.{\bf D53}, 231 (1996).
\bibitem{Eichten2003}  Eichen E J,  Hill C T  and  Quigg C, Phys. Rev. Lett.{\bf 71}, 4116 (1993).
\bibitem{Bab03} BaBar Collaboration, Aubert B {\it et al.},
            Phys. Rev. Lett. {\bf 90}, 242001 (2003).

\bibitem{Cle03} CLEO Collaboration,  Besson D {\it et al.},
            Phys. Rev. D {\bf 68}, 032002 (2003).

\bibitem{Bel04} Belle Collaboration,  Mikani Y {\it et al.},
            Phys. Rev. Lett. {\bf 92}, 012002 (2004).
\bibitem{God91} Godfrey S and  Kokoski R,
            Phys. Rev. D {\bf 43}, 1679 (1991).
         Ebert D , Galkin V O, and Faustov R N,
            Phys. Rev. D {\bf 57}, 5663 (1998).
       Pierro  M Di and Eichten E,
            Phys. Rev. D {\bf 64}, 114004 (2001).
        Lucha W and  Sch\"oberl F,
            Mod. Phys. Lett A {\bf 18}, 2837 (2003).
        Godfrey S,
            J. Phys. Conf. Ser. {\bf 9}, 59 (2005).
\bibitem{Brambilla2007} Brambilla N , in Proceedings of the VIII$^{th}$ International Workshop
on Heavy Quarks and Leptons (HQL06), Munich (2006), edited by
S.Recksiegel et al., eConf C0610161, 51 (2007) [hep-ph/0702105v2].
\bibitem{Brambilla2002}Brambilla  N, Sumino Y  and  Vairo A, Phys.
Rev. {\bf D65}, 034001 (2002).
\bibitem{AjayRai2008} Rai A K, Patel B and  Vinodkumar P C, Phys. Rev. {\bf C78}, 055202 (2008).
\bibitem{Ajay2008} Rai A K, Pandya  J N and  Vinodkumar P C, Eur. Phys. J. {\bf A4}, 77 (2008).
\bibitem{JNPandya2001}  Pandya J  N and  Vinodkumar P C, Pramana J. Phys {\bf57}, 821 (2001).
\bibitem{Radford2007}Radford S F and Repko W W, Phys. Rev. {\bf D75}, 074031 (2007).
\bibitem{BuchmullerTye1981} Buchm$\ddot{u}$ller and Tye, Phys. Rev. {\bf D24}, 132 (1981).
\bibitem{Martin1980}  Martin A, Phys. Lett. {\bf B93}, 338 (1980).
\bibitem{Amartin1979}  Martin A, Phys. Lett. {\bf B82}, 272 (1979).
\bibitem{Quiggrosner1977}  Quigg C and  Rosner J L, Phys. Lett. {\bf B71}, 153 (1977).
\bibitem{QuiggRosner1979}  Quigg  C and Rosner J L, Phys. Rep. {\bf 56}, 167 (1979).
\bibitem{Eichten1978} Eichten E \emph{et al.}, Phys. Rev. {\bf D17}, 3090 (1978).
\bibitem{vijayakumar2004} VijayaKumar K B, Hanumaiah B and Pepin S, Eur.Phys. J. {\bf A19}, 247 (2004).
\bibitem{Altarelli1982} Altarelli G, Cabibbo N, Corbo G, Maiani L and Martinelli G, Nucl. Phys. {\bf B208}, 365 (1982).
\bibitem{Ebert2003} Ebert D, Faustov R N and Galkin V O, Phys. Rev. {\bf D67}, 014027 (2003).
\bibitem{Lakhina2006} Lakhina O and  Swanson E S,  Phys. Rev. {\bf D74}, 014012 (2006).
\bibitem{Choi2007} Choi H M,  Phys. Rev. {\bf D75}, 073016 (2007).
\bibitem{Gunnar2000} Bali G S \emph{et al.}, Phys. Rev {\bf D62}, 054503 (2000);Bali G S, Phys. Rep. {\bf 343}, 1 (2001).
\bibitem{Alexandrou2003}  Alexandrou C , P de Forcrand and John O,
Nucl. Phys. Proc. Suppl {\bf119}, 667 (2003).

\bibitem{Badalian2002} Badalian A M ,  Bakker B L G and  Simonov Yu  A, Phys. Rev. {\bf D66}, 034026 (2002).
\bibitem{Badalian2008}  Badalian A M ,  Bakker B L G  and Danilkin I V  ,  arXiv:hep-ph/0805.2291.v1.
\bibitem{Albertus2005} Albertus C \emph{et al.}, Phys. Rev. {\bf D71}, 113006 (2005).
\bibitem{Fabre1988} M. Fabre de la Ripelle , Phys. Lett. {\bf B205},97 (1988).



\bibitem{EbertMod6012003}  Ebert D, Faustov R N and Galkin V O, Mod. Phys. Lett {\bf A18} 601-608 (2003).
\bibitem{Lansberg2008}  Lansberg  J P and Pham T N, Phys.
Rev. D {\bf 74}, 034001 (2006) , Phys. Rev. {\bf D75}, 017501
(2007), arXiv:hep-ph/0804.2180v1.
\bibitem{Kim2005}  Kim C S, Lee T and  Wang G L, Phys. Lett. {\bf B606}, 323(2005),arXiv:hep-ph/0411075.
\bibitem{Rosner2006} Rosner  J L et al. (CLEO Collaboration), Phys. Rev. Lett. {\bf 96}, 092003 (2006).
\bibitem{RaiPhD2005}   Rai A K (PhD Thesis), Sardar Patel University (2005)
\bibitem{Bhavin2009} Patel B and  Vinodkumar P C, J. Phy. G : Nucl Part. Phy. \textbf{36}, 035003 (2009).
\bibitem{AKRai2002}  Rai A K,  Parmar R H and  Vinodkumar P C, J. Phys. G: Nucl. Part. Phys.
 {\bf 28}, 2275(2002).
\bibitem{AKRai2006} Rai A K  and Vinodkumar P C, Pramana J. Phys. {\bf 66}, 953 (2006).
\bibitem{Branes2005} Branes T, Godfrey S and Swanson E S, Phys.Rev. {\bf D72}, 054026 (2005).
\bibitem{Voloshin2008}  Voloshin M B, Prog. Part. Nucl. Phys. \textbf{61}, 455 (2008);arXiv:hep-ph/0711.4556v3.
\bibitem{Eichten2008} Eichten E, Godfrey S, Mahlke H and Rosner J L, Rev. Mod. Phys. \textbf{80}, 1161
(2008).
\bibitem{Gerstein1995} Gershtein S S, Kiselev V V, Likhoded A K and Tkabladze A V, Phys. Rev. {\bf D51}, 3613 (1995).
\bibitem{PDG2008} Amsler C \emph{et al.} (Particle Data Group), Phys. Lett. \textbf{B}, 1 (2008)
\bibitem{Lucha1999}  Lucha W and  Sch\"{o}berl  F, Int. J. Mod. Phys. {\bf C10}, 607 (1999), arXiv:hep-ph/9811453v2.
\bibitem{Lahde2000} T A L$\ddot{a}$hde, C J Nyf$\ddot{a}$lt and D O Riska, Nucl. Phys.{\bf A674}, 141
(2000).
\bibitem{Quang Ho-Kim1998}   Quang Ho-Kim and Pham Xuan-Yem, ``The particles and their
interactions: Concept and Phenomena" Spinger-Verlag (1998).
\bibitem{Vanroyenaweissskopf} Van Royen R and  Weisskopf V F, Nuovo Cimento {\bf50}, (1967).
\bibitem{Hwang1997}  Hwang D S and Gwang-Hee Kim, Z. Phys. {\bf C76}, 107 (1997).
\bibitem{Wang2006} Wang G L, Phys. Lett. {\bf B633}, 492 (2006).
\bibitem{Ebert2006}  Ebert D \emph{et al.}, Phys. Lett. {\bf B634}, 214 (2006).
\bibitem{Cvetic2004} Cvetic G \emph{et al.}, Phys. Lett. {\bf B596}, 84 (2004).
\bibitem{Bodwin1995}   Bodwin G T, Lee J and Sinclair D K, Phys. Rev. {\bf D51}, 1125 (1995).
\bibitem{Berezhnoy1996} Berezhnoy A V, Kiselev V V and Likhoded A K, Z. Physik {\bf A336}, 89 (1996).
\bibitem{EBraaten1995} Braaten E and Fleming S, Phys. Rev. {\bf D52}, 181 (1995).
\bibitem{Gvetic2004}Gvetic G\emph{et al.}, Phys. Lett{\bf B596}, 84
(2004).
\bibitem{Narison2001}  Narison S, Phys. Rev. Lett.{\bf B520}, 115 (2001).
\bibitem{Follana2007}  Follana E \emph{et al.}, (HPQCD and UKQCD Collabs.),
[arXiv:0706.1726](2007).
\bibitem{Aubin2005}  Aubin C \emph{et al.}, Phys. Rev. Lett.{\bf 95}, 122002 (2005).

\bibitem{Khan2007} Khan A A\emph{et al.}, (QCDSF Collaboration),
Phys. Lett. \textbf{B652}, 150 (2007).
\bibitem{Chiu2005}  Chiu  T W \emph{et al.}, Phys. Lett. \textbf{B624},
31 (2005).
\bibitem{Rosner2008} Rosner J L and Stone S, arXiv:hep-ex/0802.1043v1.
\bibitem{Gerstein1998} Gerstein S S \emph{et al.}, arXiv:hep-ph/9803433v1.
\bibitem{Gottfried1978}  Gottfried K, Phys. Rev. Lett. {\bf40}, 598 (1978).
\bibitem{Yosef1989} Yosef NIR, Phys. Lett. {\bf B221}, 184 (1989).
\bibitem{Cabibbo1978} N Cabibbo \emph{et al.}, Phys. Lett. {\bf B79}, 109 (1978).
\bibitem{Luke1993} Michael Luke and Martin J. Savage,
arXiv:hep-ph/9308287v2.
\bibitem{Voloshin1979} Voloshin M B, Nucl. Phys. {\bf B154}, 365 (1979).
\bibitem{Kuang1990} Kuang Y P and Yan T M , Phys. Rev. {\bf D41}, 155 (1990).

\bibitem{Bodwin2008} Bodwin G T \emph{et. al}, Phys. Rev. {\bf D77}, 094017 (2008).
\bibitem{Juan-Luis2008} Juan-Luis Domenesh-Garret and Miguel-Angel Sanchis-Lozano, Comput. Phys. Commun. {\bf 180}, 768 (2009);arXiv:hep-ph/0805.2916v3,arXiv:hep-ph/0805.2704v1.

\bibitem{Stefano2007}Villa S, arXiv:hep-ex/0707.0263v1.
\bibitem{Link:2003bd} Link J M {\it et al.} [FOCUS Collaboration],Phys.\ Lett. {\bf B586}, 11 (2004).
\bibitem{Abe:2006xm} Abe K{\it et al.} [Belle Collaboration], Belle Report BELLE-CONF-0643, hep-ex/0608031.
\bibitem{Aubert:2006mh} Aubert B{\it et al.} [BaBar Collaboration], Phys. Rev. Lett. {\bf97}, 222001 (2006).
\bibitem{Artuso:2005ym} Artuso M{\it et al.} [CLEO Collaboration], Phys. Rev. Lett. {\bf95}, 251801 (2005).
\bibitem{Aubin:2005ar} Aubin C{\it et al.} Phys. Rev. Lett. {\bf95}, 122002 (2006).








\end{thebibliography}
\end{document}